\documentclass[fleqn,twoside]{article}
\usepackage{espcrc2}

\usepackage{epsfig}
\newcommand{\AmS}{{\protect\the\textfont2
  A\kern-.1667em\lower.5ex\hbox{M}\kern-.125emS}}

\newcommand{\epjc}      [3]  {{{\rm Eur.\ Phys.\ J.}\ C{#1}, {#2} ({#3})}}

\newcommand{\nima}      [3]  {{{\rm Nucl.\ Instr.\ Methods A} {{#1}, {#2} ({#3})}}}

\newcommand{\npb}       [3]  {{{\rm Nucl.\ Phys.}\ B{#1}, {#2} ({#3})}}

\def\bb{b\bar{b}}

\def\bd{B^0_d} 
\def\bs{B^0_s}
\def\bzero{B^0} 
\def\bzerob{\overline{B^0}} 
 
\def\bsb{\overline{B^0_s}}

\def\bdmix{B_d^0 \mbox{--} \overline{B_d^0}}
\def\bsmix{B_s^0 \mbox{--} \overline{B_s^0}}
\def\dmd{\Delta m_d}
\def\dms{\Delta m_s}
\def\ips{\mbox{ps}^{-1}}
\def\ups4s{\Upsilon_{4S}}
\def\vtd{V_{td}}
\def\vts{V_{ts}}
\def\Zbb{Z^0 \rightarrow b\,{\overline b}}

\title{$\bs$ Oscillation Results}

\author{St\'ephane Willocq\address[UMASS]{Department of Physics, 
        University of Massachusetts, \\ 
        Amherst, MA 01003, USA}%
}
       
\begin{document}

\begin{abstract}
We review new studies of the time dependence of $\bsmix$ mixing
by the ALEPH, DELPHI and SLD Collaborations,
with an emphasis on the different analysis methods used.
Combining all available results yields a preliminary lower limit
on the oscillation frequency of $\dms > 14.4$ $\ips$ at the 95\% C.L.
\vspace{1pc}
\end{abstract}

\maketitle

\section{INTRODUCTION}

  Studies of the time dependence of $\bdmix$ and $\bsmix$ mixing
continue to play an important role in the exploration of both the
Cabibbo-Kobayashi-Maskawa (CKM) quark mixing matrix
and the phenomenon of CP violation.
The $\bd$ oscillation frequency is related to the poorly
known CKM element $\vtd$ according to~\cite{Buras}
\begin{eqnarray}
  \dmd & = &
       \frac{G_F^2}{6\pi^2} m_{B_d} m_t^2 F(m_t^2 / m_W^2) f_{B_d}^2 B_{B_d}
   \nonumber \\
       &   &  \times~\eta_{QCD} \left| V_{tb}^\ast V_{td} \right|^2
  \label{eq_dmd}
\end{eqnarray}
and has recently been precisely measured by the BaBar and Belle
collaborations---the current world average is
$\dmd = 0.503 \pm 0.006$ $\ips$~\cite{BOSCWG}.
However, this measurement cannot be translated into a precise determination
of $|\vtd|$ due the 15-20\% theoretical uncertainty in
$f_{B_d} \sqrt{B_{B_d}}$ (see Ref.~\cite{Lellouch} for a review of Lattice
QCD calculations).
Uncertainties are reduced for the ratio
\begin{equation}
  \frac{\dms}{\dmd} = \frac{m_{B_s}}{m_{B_d}} \xi^2 \left|\frac{V_{ts}}{V_{td}}\right|^2 ,
\end{equation}
where the quantity $\xi \equiv (f_{B_s} \sqrt{B_{B_s}}) / (f_{B_d} \sqrt{B_{B_d}})$
is estimated to be $1.18 \pm 0.04 ^{+0.12}_{-0.00}$
from Lattice QCD~\cite{Lellouch}.
This implies that the ratio $|V_{ts}/V_{td}|$ can be determined
with an uncertainty smaller than that for $|V_{td}|$~\cite{Kronfeld}.
In the Wolfenstein parameterization of the CKM matrix, we have
$\dmd \propto |\vtd|^2 \simeq A^2 \lambda^6 [(1 - \rho)^2 + \eta^2]$
and $\dms \propto |\vts|^2 \simeq A^2 \lambda^4$,
where $\lambda = 0.2237 \pm 0.0033$ and $A = 0.819 \pm 0.040$~\cite{Ciuchini},
but $\rho$ and $\eta$ are not well known.
Studies of $\bd$ and $\bs$ oscillations thus provide some of the
strongest constraints on the CKM unitarity triangle parameters $\rho$
and $\eta$.

  Experimental studies of $\bs$ oscillations require two main ingredients:
(i) reconstruction of the $\bs$ decay and its proper time,
(ii) determination of the $\bs$ or $\bsb$ flavor at both production and
decay to classify the decay as either `mixed' (if the tags disagree)
or `unmixed' (otherwise).
The significance for a $\bs$ oscillation signal can be approximated
by~\cite{Moser}
\begin{equation}
  S = \sqrt{\frac{N}{2}}\: f_s\: \left[1 - 2\, w\right]\:
      e^{-\frac{1}{2} (\dms \sigma_t)^2} ,
  \label{eq_signif}
\end{equation}
where $N$ is the total number of decays selected,
$f_s$ is the fraction of $\bs$ mesons in the selected sample,
$w$ is the probability to incorrectly tag a decay as mixed or unmixed
(i.e. the mistag rate)
and $\sigma_t$ is the proper time resolution.
The proper time resolution depends on both the decay length resolution
$\sigma_L$ and the momentum resolution $\sigma_p$ according to
$\sigma_t^2 = (\sigma_L m_B/ p)^2 + (t\, \sigma_p/p)^2$.
Based on the Wolfenstein parameterization, we see that
$\dms / \dmd \propto 1 / \lambda^2$, which is of order of 20
(the other Wolfenstein parameters are of order 1).
Therefore, $\bs$ oscillations are expected to be much more rapid
than $\bd$ oscillations.
The ability to resolve such rapid oscillations thus requires excellent
decay length and momentum resolution, and benefits from having
a low mistag rate and a high $\bs$ purity.

\section{RECONSTRUCTION METHODS}

  The study of the time dependence of $\bsmix$ mixing
has been performed with different analysis techniques,
ranging from fully inclusive to fully exclusive reconstruction of $\bs$ decay
candidates.
The study of $\bs$ oscillations is more chalenging than that of
$\bd$ oscillations due to two main differences.
First, only about 10\% of $b$ quarks fragment into $\bs$ mesons, as compared
to about 40\% into $\bd$ mesons. Second, the $\bs$ oscillation frequency
is expected to be at least a factor of 20 larger than that for $\bd$ oscillations.
To address this, sophisticated analyses have been
developed with an emphasis on lowering the mistag rate, increasing
the $\bs$ purity and, especially, improving the proper time resolution,
all of which affect the sensitivity to $\bs$ oscillations.

  The production flavor tag combines a number of different individual tags.
The single most powerful tag exploits the large polarized forward-backward
asymmetry in $\Zbb$ decays. This tag is available at SLD thanks to the
large electron beam polarization ($P_e = 73\%$).
A left- (right-) handed incident electron tags the quark produced
in the forward hemisphere as a $b$ ($\overline{b}$) quark.
This method yields a mistag rate of $~28\%$ with nearly 100\% efficiency.
Tags used in all analyses rely on charge information from the event hemisphere
opposite that of the $\bs$ candidate:
(i) charge of lepton from the direct transition $b \to \ell^-$,
(ii) momentum-weighted jet charge,
(iii) secondary vertex charge,
(iv) charge of secondary vertex kaon from the dominant transition $b \to c \to s$,
(v) charge dipole of secondary vertex (SLD only).
Other tags from the same hemisphere as the $\bs$ candidate are also used:
(i) unweighted (or weighted) jet charge, and (ii) charge of fragmentation kaon
accompanying the $\bs$ meson.
These various tags are combined on an event-by-event basis to yield an
average mistag rate of approximately 22\% at SLD and 27-29\% at LEP.

  The analyses differ in the way the $\bs$ decay is reconstructed and thus in the
way the decay flavor is determined. Three general classes can be identified:
inclusive, semi-exclusive and fully exclusive. Inclusive analyses benefit
from the large available statistics but suffer from low $\bs$ purity,
whereas more exclusive analyses benefit from higher purity and resolution
but suffer from the lack of statistics
(this is particularly true for the fully exclusive analyses).
Several analyses are discussed below to highlight these differences.

  Inclusive analyses have been performed by ALEPH, DELPHI, OPAL and SLD.
The SLD charge dipole analysis is the most sensitive fully inclusive
method~\cite{SLDdipole}.
It aims to reconstruct the $b$-hadron decay chain topology.
This method takes full advantage of the superb decay length resolution
of the SLD CCD pixel vertex detector to separate secondary tracks (from the $B$
decay point) from tertiary tracks (from the $D$ decay point).
The decay length resolution is parameterized by the sum of two Gaussians with
$\sigma_L = 78\:\mu$m (60\% fraction) and $304\:\mu$m (40\%),
whereas the momentum resolution is parameterized with
$\sigma_p / p = 0.07$ (60\%) and 0.21 (40\%).
A ``charge dipole'' $\delta Q$ is defined as the distance between
secondary and tertiary
vertices signed by the charge difference between them such that
$\delta Q > 0$ ($\delta Q < 0$) tags $\bzerob$ ($\bzero$) decays.
The average decay flavor mistag rate is estimated to be 22\% and is
mostly due to decays producing two charmed hadrons.
A sample of 11,462 decays is selected with a $\bs$ purity estimated to be 16\%
(higher than the production rate of 10\% due to the fact that
only neutral decays are selected).

  The most sensitive of all analyses is the ALEPH inclusive lepton
analysis~\cite{ALEPH}, which selects semileptonic $B$ decays.
In this analysis, a $D$ meson is reconstructed inclusively based
on topological and kinematical properties of the decay, and
a resultant $D$ track is vertexed with the lepton and the $b$-hadron
direction (from the jet direction) to form a $B$ decay vertex.
The average decay length and momentum resolutions are
$\sigma_L = 251\:\mu$m (75\% fraction) and $718\:\mu$m (25\%),
$\sigma_p / p = 0.064$ (60\%) and 0.20 (40\%).
Fairly loose selection criteria are used at the various stages
of the analysis to obtain a high statistics sample of 74,026 events.
The analysis relies on several neural network algorithms to perform
the following tasks:
production flavor tagging, $\bb$ event selection, direct $(b \to \ell)$
lepton selection, and $\bs$ fraction enhancement.
To maximize sensitivity to $\bs$ oscillations, the analysis
incorporates all the information event by event, including estimates
of the decay length and momentum resolution.

  Semi-exclusive analyses have been performed by ALEPH,
CDF, DELPHI, OPAL and SLD.
$\bs$ decays are partially reconstructed in the modes
$\bs \to D_s^- \ell^+ \nu_\ell X$ and
$\bs \to D_s^- h^+ X$,
where $h$ represents any charged hadron (or system of several hadrons)
and the $D_s^-$ meson decay is either fully or partially reconstructed
in the modes $D_s^- \to \phi\pi^-$, $K^{\ast 0} K^-$, $K^0 K^-$,
$\phi\pi^-\pi^+\pi^-$, $\phi \ell^- \overline{\nu_\ell}$, etc.

  The most sensitive semi-exclusive analysis performed by DELPHI selects
436 $D_s^- \ell^+$ events~\cite{DELPHI}.
The small statistics is compensated by the
high $\bs \to D_s^- \ell^+\nu_\ell$ purity, estimated to be $\sim 53\%$,
and the good decay length and momentum resolution,
$\sigma_L = 200\:\mu$m (82\% fraction) and $670\:\mu$m (16\%),
$\sigma_p / p = 0.07$ (82\%) and 0.16 (16\%).
Analyses selecting $D_s^- h^+$ final states benefit from
higher statistics but are less sensitive than those selecting
$D_s^- \ell^+$ states because of lower $\bs$ purity and worse proper time
resolution.
The SLD $D_s$+Tracks analysis~\cite{SLDDstracks}
combines fully reconstructed $D_s$
mesons with either a lepton or one (or more) charged hadron(s).
It contributes especially at large values of $\dms$ thanks to
a $\bs$ purity of 40\% and the
best available decay length resolution:
$\sigma_L = 50\:\mu$m (60\% fraction) and $151\:\mu$m (40\%).

  Finally, fully exclusive analyses have been performed by ALEPH~\cite{ALEPH}
and DELPHI~\cite{DELPHIexcl} via the (all charged particles) modes
$\bs \to D_s^-\pi^+$, $D_s^- a_1^+$, $\overline{D^0} K^- \pi^+$,
and $\overline{D^0} K^- a_1^+$ (last two for DELPHI only),
where the $D_s^-$ and $\overline{D^0}$ are fully reconstructed.
The decays $\bs \to D_s^{\ast-}\pi^+$, $D_s^{\ast-}a_1^+$ and
$D_s^{(\ast)-}\rho^+$ are also reconstructed by adding one or more
photons to the above final states (ALEPH only) or by including the
``satellite'' mass region below the $\bs$ mass peak.
The number of decay candidates is 80 for ALEPH and 44 for DELPHI with signal
purities of approximately 36\% and 50\%, respectively.
The main advantage of the exclusive method is its excellent
proper time resolution
with a negligible contribution from momentum resolution ($\sim 0.5\%$).
As a result, unlike all other methods,
$\sigma_t$ does not grow significantly with increasing proper time $t$ and thus
the oscillation amplitude is not damped as $t$ increases.
Due to limited statistics, this method is not competitive with respect
to the inclusive and semi-exclusive methods.
However, this is the method of choice for future studies of $\bs$ oscillations
at hadron colliders.

\section{RESULTS}

\begin{figure}
\begin{center}
  \epsfxsize8.5cm
  \vspace{-0.9cm}
  \epsfbox{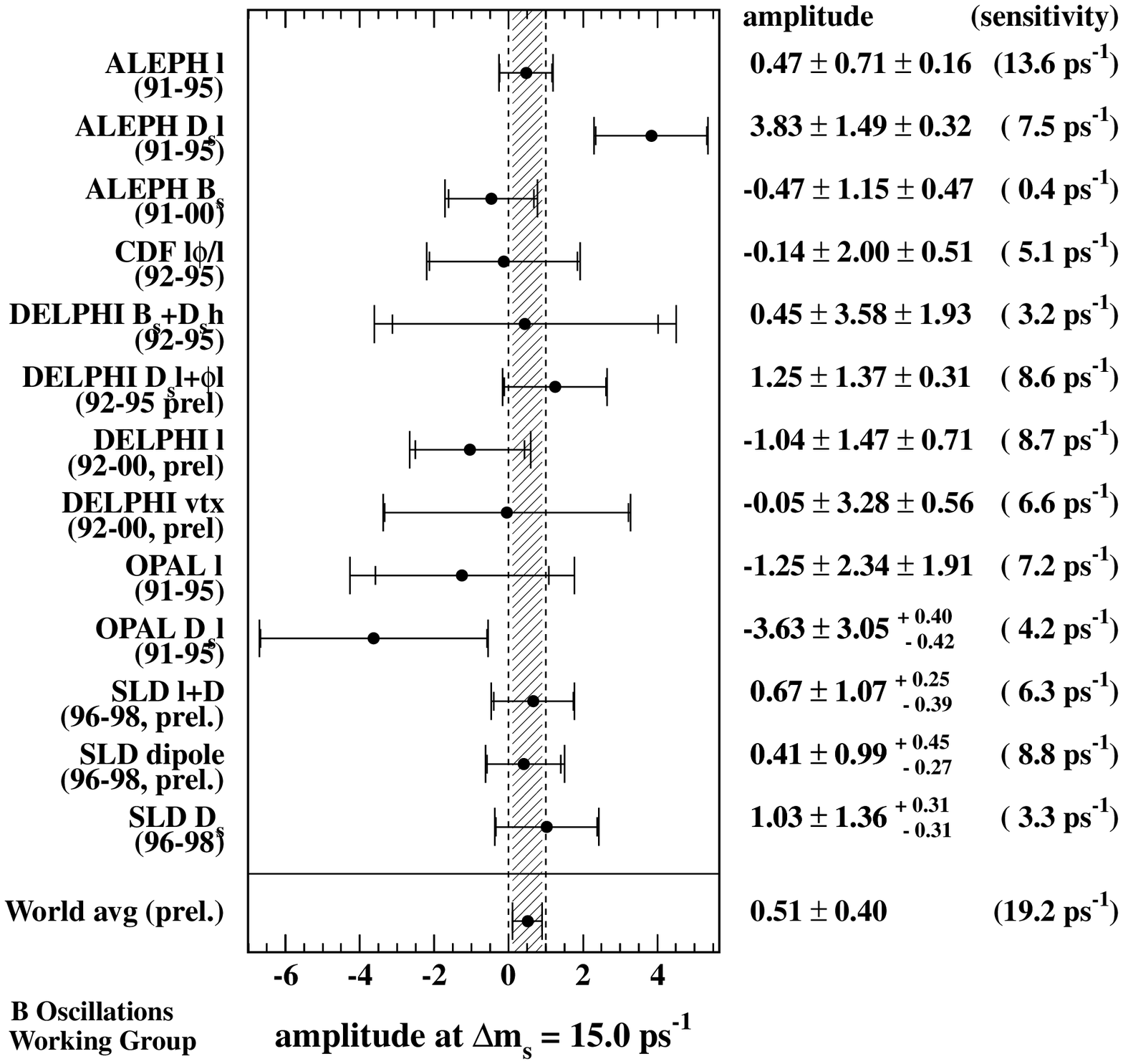}
\end{center}
\vspace{-1.0cm}
\caption{Measurements of the $\bs$ oscillation amplitude
  for $\dms = 15~\ips$. }
\label{fig_ampW}
\end{figure}

  Studies of the time dependence of $\bsmix$ mixing are carried out
with the amplitude method, which is equivalent to a normalized Fourier
transform~\cite{Moser}.
The oscillation amplitude $A$ is expected to be $A=0$ ($A = 1$)
for frequencies sufficiently far from (close to) the true
oscillation frequency.
All available measurements of the oscillation amplitude at $\dms = 15~\ips$ are
summarized in Figure~\ref{fig_ampW}. Also shown are the sensitivities for each
analysis to set a 95\% C.L. lower limit on $\dms$ ($\dms$ value at which
$1.645\:\sigma_A = 1$).
%  A comparison between the sensitivities of the different methods is shown
%in Fig.~\ref{fig_methods}, which clearly shows that the inclusive methods
%dominate.
%\begin{figure}
%\begin{center}
%  \epsfxsize8.3cm
%  \vspace{-0.7cm}
%  \epsfbox{sensit_methods.eps}
%\end{center}
%\vspace{-1.0cm}
%\caption{Sensitivities for the different methods as a function
%  of $\dms$. }
%\label{fig_methods}
%\end{figure}

\begin{figure}
\begin{center}
  \epsfxsize8.3cm
  \vspace{-0.7cm}
  \epsfbox{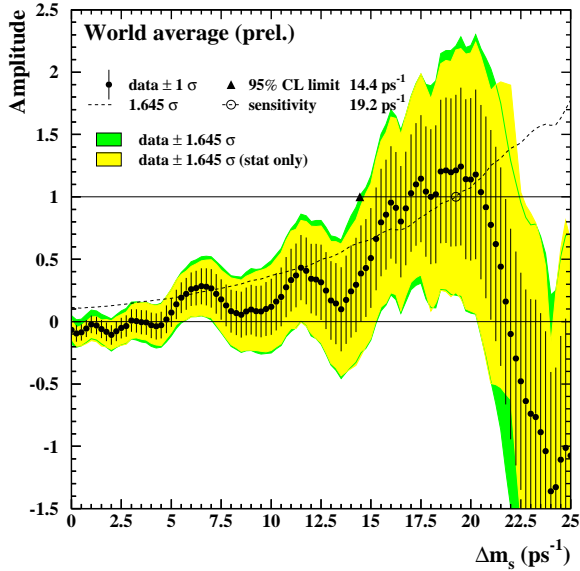}
\end{center}
\vspace{-1.0cm}
\caption{World average $\bs$ oscillation amplitude as a function of
         $\dms$. Values of the frequency for which $A + 1.645\:\sigma_A < 1$
         are excluded at the 95\% C.L.}
\label{fig_afitW}
\end{figure}

  The measured oscillation amplitudes are combined~\cite{BOSCWG},
taking statistical and systematic correlations into account,
to obtain the world average amplitude spectrum
shown in Figure~\ref{fig_afitW}.
The combination also corrects for the different input parameters
used ($\dmd$, $b$-hadron lifetimes and production rates).
Furthermore, the amplitude statistical uncertainties
measured by the inclusive analyses are adjusted to take into
account the $\bs$ production rate of $(9.3 \pm 1.1)\%$.
The rise in statistical error as $\dms$ increases comes from the fact
that an increasingly smaller fraction of the data sample has sufficient
proper time resolution to resolve more rapid oscillations;
the better the resolution, the smaller the rise.
The preliminary combined amplitude spectrum excludes mixing $(A = 1)$ for
$\dms < 14.4$ $\ips$ at the 95\% C.L., whereas the sensitivity is 19.2~$\ips$.
The significance of the deviation from $A = 0$ near $\dms = 17.5~\ips$
is $2.3\:\sigma$.
It is interesting to note that, while the sensitivity has been increasing
steadily over the past 8 years, the limit has remained near 15~$\ips$
for the past 3 years. In August 1999, the limit was $\dms > 14.3$~$\ips$
but the sensitivity was only 14.7~$\ips$.

  Many of the LEP and SLD analyses have been or are being finalized.
One will thus have to wait for the next generation of experiments
to measure the $\bs$ oscillation frequency.
Prospects for such a measurement during Run 2 of the Fermilab
Tevatron are excellent.

  I wish to thank D.Abbaneo, P.Kluit, F.Parodi, O.Schneider,
and A.Stocchi for their help with the preparation of the talk.

%**************************

\end{document}